# A physical mechanism of the generation of stable positive kinetic energy systems and a qualitative explanation of the proportions of the four ingredients in the universe

Huai-Yu Wang

Department of Physics, Tsinghua University, Beijing, 100084 China
wanghuaiyu@mail.tsinghua.edu.cn

**Abstract:** The author's opinion is that the negative energy solutions of the Dirac equation mean that a particle can be of negative kinetic energy (NKE) besides positive kinetic energy (PKE). We think that NKE particles are dark ones and NKE matter is dark matter. In our previous works, the dark matter theory of the NKE version and the dark energy theory that matched dark matter theory were put forth. In this work, we investigate the topics related to the metamorphosis of objects between PKE and NKE. We first evaluate the collisions between a PKE and a NKE particles. A scenario of accelerating PKE particles is raised. We put forth the cosmic dark radiation background and gravity potential background. In the universe, negative energy is predominating. In the observable universe, substances constitute stable PKE systems. The total energy of every such system is negative. We propose a mechanism that NKE substances combine into stable PKE systems. Macroscopically, NKE objects can constitute stable PKE astrophysical systems by means of gravity between them. Microscopically, NKE particles can combine into stable PKE systems by means of attractive interactions between them, say, Coulomb attraction. Currently, people think that there are four ingredients in the universe: photons $\Omega_{R0}$, matter $\Omega_{M0}$, dark matter $\Omega_{DM}$, and dark energy $\Omega_{\Lambda 0}$. We analyze the order of the appearance of the four ingredients and conclude that qualitatively, their proportions should be $\Omega_{\Lambda 0} > \Omega_{DM0} > \Omega_{M0} > \Omega_{R0}$.

**Résumé** : L'opinion de l'auteur est que les solutions d'énergie négative de l'équation de Dirac signifient qu'une particule peut avoir une énergie cinétique négative (NKE) en plus de l'énergie cinétique positive (PKE). Nous pensons que les particules à ECN sont des particules noires et que la matière à NKE est de la matière noire. Dans nos travaux précédents, nous avons proposé la théorie de la matière noire dans sa version à NKE ainsi que la théorie de l'énergie noire qui correspond à la théorie de la matière noire. Dans ce travail, nous étudions les sujets liés à la métamorphose des objets entre l'PKE et l'NKE. Nous évaluons d'abord les collisions entre une particule à PKE et une particule à NKE. Un scénario d'accélération des particules à PKE est proposé. Nous avançons l'idée du fond cosmique de rayonnement noire et du fond potentiel gravitationnel. Dans l'univers, l'énergie négative prédomine. Dans l'univers observable, les substances constituent des systèmes stables à PKE. L'énergie totale de chaque système de ce type est négative. Nous proposons un mécanisme selon lequel les substances à NKE se combinent pour former des systèmes stables à PKE. Macroscopiquement, les objets à NKE peuvent constituer des systèmes astrophysiques stables à PKE grâce à la gravité qui agit entre eux. Microscopiquement, les particules à NKE peuvent se combiner pour former des systèmes stables à PKE grâce aux interactions attractives entre elles, par exemple l'attraction de Coulomb. Actuellement, on pense qu'il y a quatre ingrédients dans l'univers : les photons $\Omega_{R0}$, la matière $\Omega_{M0}$, la matière noire $\Omega_{DM}$ et l'énergie noire $\Omega_{\Lambda 0}$. Nous analysons l'ordre d'apparition des quatre ingrédients et concluons que qualitativement, leurs proportions devraient être

$$\Omega_{\Lambda 0} > \Omega_{DM0} > \Omega_{M0} > \Omega_{R0} \ .$$

**I. INTRODUCTION**

The Dirac equation, as a relativistic quantum mechanics equation, has both positive and negative energy solutions. We think that they actually are positive kinetic energy (PKE) and negative kinetic energy (NKE) solutions.[1] This shows that a particle can have either PKE or NKE. Our opinion is that the PKE and NKE solutions should be treated on an equal footing. The low-momentum approximations of the Dirac equation for the PKE and NKE are the Schrödinger equation and NKE Schrödinger equation, respectively. The content of physical research up to now has focused on matter with PKE. We conducted a series of studies on the NKE substances.[1–12] All the physical laws we have known remain unchanged, such as thermodynamics, statistical mechanics, classical mechanics, quantum mechanics. The fundamental laws of physics are of symmetry with respect to the PKE and NKE.[5]

Our basic point of view is that the NKE substances are dark matter. When a particle has PKE, it is visible to us, while when it has NKE, it is invisible to us. Therefore, we have never seen NKE matter. In previous work, we gave the reason why the NKE matter is dark to us.[3,11]

A PKE substance can absorb photons so as to transit from a lower energy level to a higher level. It can also spontaneously transit from a higher level to a lower one, accompanied by emitting photons. The emitted photons are received by our instruments, allowing us to "see" the material that emits the photons. We say that PKE matter and energy are matched, and they both have positive temperature and positive pressure. In the following, positive energy is simply called energy or photons or radiation.

The energy spectrum of a NKE system has an upper limit but no lower limit. So it has a negative temperature.[3] The higher the energy level, the more stable the state of the system, and the lower the energy level, the more unstable the state. The highest energy level is the most stable and is the ground state. Therefore, spontaneous transitions can occur only from lower energy levels to higher ones. We believe that such spontaneous transition processes release negative energy which is dark energy. The theory of the dark energy was given.[11] Dark energy is photons with negative energies, called dark photons or negative photons. Dark photon fields are also called dark radiation or negative radiation. Its statistical mechanics and thermodynamic formulas have been presented. The energy spectrum of the dark radiation has an upper limit but no lower limit, and therefore, it has a negative temperature. When a system absorbs negative photons, an excited transition from a higher energy level to a lower energy level occurs. Thus, the NKE system and the negative radiation are matched, and they both have negative temperature and negative pressure.

It is believed that our universe is of the symmetries of PKE and NKE, of PKE matter and NKE matter, of matter and dark matter, of energy and dark energy, and of matter-radiation matching and dark matter-dark radiation matching.

The observable objects in the universe are composed of PKE particles, and the instruments we make are all composed of PKE particles. The PKE systems cannot absorb and emit dark energy. Therefore, the NKE systems and negative radiation are dark to us.

We suggested experiments to detect NKE electrons.[1,12]

A particle may move with either PKE or NKE, since either is the solution of the Dirac equation. Is it possible for a particle to transform between the PKE and NKE in a process of motion? Such a transformation is called metamorphosis. We believe that metamorphosis may occur under the premise of conservation of energy.

The single-particle Dirac equation contains both PKE and NKE energy levels. The metamorphosis of a particle is by no means the transition between the PKE and NKE levels. We think that an isolated particle will never metamorphose. The metamorphosis of a particle is closely related to the interactions with its surroundings. For example, when a particle enters a potential barrier, it metamorphoses from PKE to NKE.[2] In the present work, we are going to discuss how two NKE particles combine into one stable PKE system.

There are three types of energy spectrum.[3] Type I energy spectrum has a lower limit but no upper bound, and a system with such a spectrum is a PKE one with a positive temperature. Type II spectrum has an upper limit but no lower limit, and a system with such a spectrum is a NKE one with a negative temperature. Type III energy spectrum has both a lower and an upper bound, and the temperature can be the entire range except zero. When the system is at energy levels near the lower limit, the system has a positive temperature; while when the system is at energy levels near the upper limit, the system has a negative temperature. In such a system, a state with a negative temperature has an energy higher than that with a positive temperature.

What we are now well known is the type I energy spectrum. We have not encountered the systems with Type II energy spectrum. Some theoretical studies about the Type II systems were given.[1,3] As early as 1950s, type III energy spectrum was experimentally realized in spin systems.[13–19] Recently, this type of spectrum was achieved in a photonic crystal.[20] The concept of negative temperature has been proposed for a long time, and there has been much theoretically and experimentally work on the study of negative temperature systems.[13–33]

Let us now consider the contact of two systems with different types of energy spectrum. Suppose that a type I system and a type III system come into contact. Since the type I system can only have a positive temperature, both systems must be at the same positive temperature when equilibrium is reached. The adiabatic demagnetization is an example.[34–39]

Imagine that a type II system and a type III system come into contact. It is inferred that both systems must be at the same negative temperature when equilibrium is reached. This is because the type II system can only have a negative temperature.

When a Type I system is in contact with a Type II system, can equilibrium be achieved? If it can, what should it be? This is to be discussed. Please note that the two systems can only have positive and negative temperatures, respectively.

According to currently popular cosmological theory, there are four ingredients in the universe:[40,41] photons, denotd as $\Omega_{R0}$, matter denoted as $\Omega_{M0}$, dark matter denoted as $\Omega_{DM}$, and dark energy denoted as $\Omega_{\Lambda 0}$. If the total amount of the four ingredients is set as 1, their respective proportions are as follows.

$$\Omega_{\Lambda 0} \sim 0.7; \ \Omega_{DM0} \sim 0.25; \ \Omega_{M0} \sim 0.05; \ \Omega_{R0} \sim 5\times 10^{-5}. \tag{1}$$

That is to say, the proportion of the dark energy is the largest, that of the dark matter is the next, and that of the photons is the smallest. Why their proportions are so remains open.

We note that a striking feature of our universe is that all the matter we can observe has PKE, and almost all of them form stable systems. The total energy of a stable system we can observe is necessarily negative. Our universe, in general, is full of negative energy. In this paper, a mechanism is proposed that NKE substances can form a stable PKE system. We attempt to give a qualitative explanation for Eq. (1).

This paper is arranged as follows. In Section II, we investigate the collisions between a PKE and a NKE particles, and a way of accelerating particles is suggested. In Section III, we study what is the final state when a PKE ideal gas and a NKE ideal gas mix. In Section IV, we introduce the concepts of the cosmic dark radiation background and the gravitational potential background. In Section V, we propose a possible mechanism for the formation of PKE stabilization systems in the universe. Section VI gives a qualitative explanation of Eq. ( 1 ). Section VII is our conclusion.

## II. TWO-PARTICLE ELASTIC COLLISION

Nowadays, the experimental search for dark matter is to let the expected dark particles collide with the particles in the instrument.[42–46] The preconceived dark particles are thought to have PKE. Some dark matter candidates were excluded.[47]

We suggested using photons or electrons to collide with tunneling electrons to verify that the latter had NKE.[1,12]

Let us consider the general problem of two-particle elastic collisions. The topic of the elastic collision between two PKE particles has been well studied.[48,49] If both particles participating in the elastic collision have NKE, then, the analysis of the elastic collision is exactly the same as that of two PKE particles since all the kinetic energies of the two particles have a minus sign before and after the collision. We are mainly concerned with the collision between a PKE and a NKE particles. Nevertheless, for comparison, we still review the collision formula for two PKE particles below. In the following equation, let the Planck constant and light speed to be unit: $\hbar = c = 1$.

Let the masses of the two particles be $m_1$ and $m_2$, respectively. Before the collision, the particle with a mass $m_1$ is the incident one and that with a mass $m_2$ is at rest and called target particle. Both energy and momentum after a collision are marked by a prime. We first discuss the case of low-momentum motion, then the case of relativistic motion, and at last the case that the incident particle has a zero rest mass.

**A. Low-momentum motion**

A.1 Both $m_1$ and $m_2$ are of PKE

We first recall the collsion of two PKE particles, as shown in Fig. 1. The conservation of the total kinetic energy is

$$\frac{\boldsymbol{p}_1^2}{2m_1} = \frac{\boldsymbol{p}_1'^2}{2m_1} + \frac{\boldsymbol{p}_2'^2}{2m_2}. \tag{2}$$

The momentum conservation gives

$$|\boldsymbol{p}_1| = |\boldsymbol{p}_1'|\cos\theta + |\boldsymbol{p}_2'|\cos\varphi \tag{3}$$

and

$$|\boldsymbol{p}_1'|\sin\theta = |\boldsymbol{p}_2'|\sin\varphi. \tag{4}$$

It is solved that

$$|\boldsymbol{p}_1'| = \frac{|\boldsymbol{p}_1|}{m_1 + m_2}[m_1\cos\theta + g(\theta)] \tag{5}$$

and

$$|\boldsymbol{p}_2'| = \frac{2m_2|\boldsymbol{p}_1|\cos\varphi}{m_1 + m_2}, \tag{6}$$

where

$$g(\theta) = \pm\sqrt{m_2^2 - m_1^2\sin^2\theta}. \tag{7}$$

The selection of the sign in Eq. (7) depends on that which of the $m_1$ and $m_2$ is greater. It is required from Eq. (5) that

$$m_1 \cos\theta \pm \sqrt{m_2^2 - m_1^2 \sin^2\theta} > 0. \tag{8}$$

We distinguish the following two cases.

$$m_1 < m_2, g(\theta) = \sqrt{m_2^2 - m_1^2 \sin^2\theta}. \tag{9}$$

$$m_1 > m_2, g(\theta) = -\sqrt{m_2^2 - m_1^2 \sin^2\theta}, 0 \leq \theta < \sin^{-1}(m_2/m_1). \tag{10}$$

It follows from Eq. (6) that the scattering angle of the $m_2$ is necessarily within the first quadrant, $0 < \varphi \leq \pi/2$.

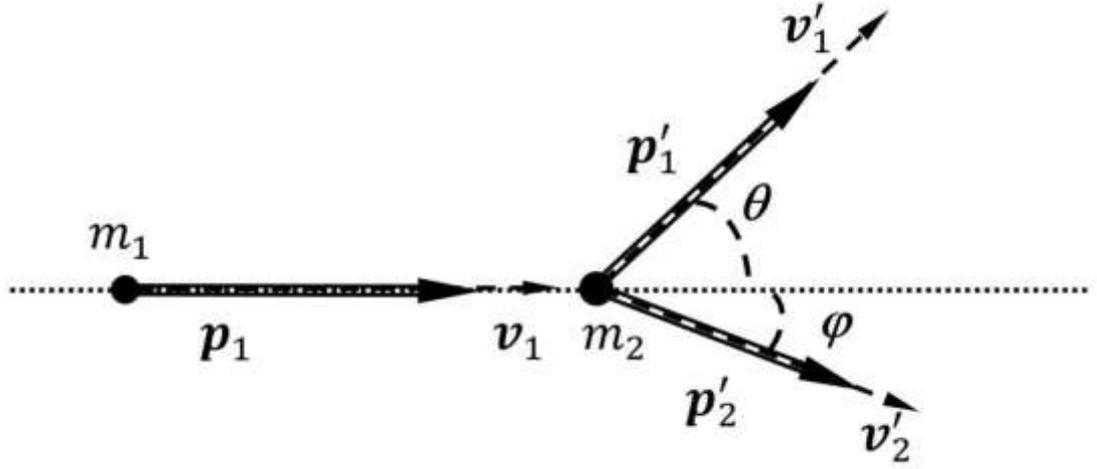

FIG. 1. An incident PKE particle with mass $m_1$ collides with a static PKE particle with $m_2$. Solid circles are the initial positions of the two particles. Dashed lines mean velocities. The double-solid lines are momenta. When $m_1 = 0$, the incident particle is a photon.

A.2 $m_1$ is of PKE and $m_2$ is of NKE

The target particle is of NKE, see Fig. 2. The conservation of the total kinetic energy is

$$\frac{p_1^2}{2m_1} = \frac{p_1'^2}{2m_1} - \frac{p_2'^2}{2m_2}. \tag{11}$$

The conservation of the total momentum is still Eqs. (3) and (4). It is then solved that

$$|p_1'| = \frac{|p_1|}{m_1 - m_2}[m_1 \cos\theta + g(\theta)] \tag{12}$$

and

$$|p_2'| = \frac{2m_2 |p_1| \cos\varphi}{m_2 - m_1}. \tag{13}$$

From these two equations, following two cases are distinguished.

$$m_1 < m_2, \quad g(\theta) = -\sqrt{m_2^2 - m_1^2 \sin^2\theta}, \quad 0 \leq \varphi < \pi/2. \tag{14}$$

$$m_1 > m_2, \quad g(\theta) = -\sqrt{m_2^2 - m_1^2 \sin^2\theta}, \quad 0 \leq \theta < \sin^{-1}(m_2/m_1), \quad \pi/2 \leq \varphi < \pi. \tag{15}$$

In both cases, $g(\theta)$ must take the minus sign.

Although the NKE particle cannot be directly detected, according to the formulas above, we accurately know the NKE particle's momentum $p_2'$ and velocity $v_2'$ after the collision as shown in Fig. 2. Such a NKE particle can be utilized to collide with a PKE particle.

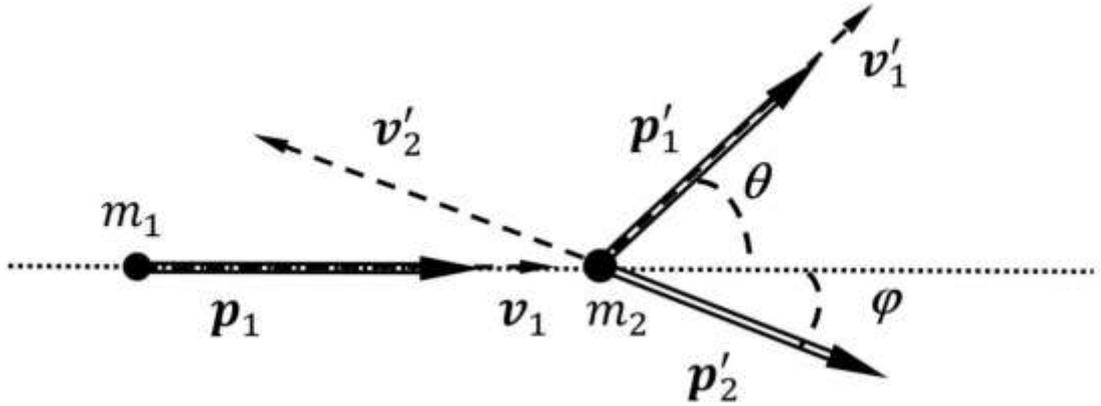

FIG. 2. An incident PKE particle with mass $m_1$ collides with a static NKE particle with $m_2$. Solid circles are the initial positions of the two particles. Dashed lines mean velocities. The double-solid lines are momenta. Please note that the directions of the velocity and momentum of a NKE particle are opposite to each other. When $m_1 = 0$, the incident particle is a photon.

A.3 $m_1$ is of NKE and $m_2$ is of PKE

The incident particle is of NKE and the target particle is of PKE, as shown in Fig. 3. Before the collision, the total momentum points to the left, and the direction remains after the collision. The conservation of the total kinetic energy is

$$-\frac{p_1^2}{2m_1} = -\frac{p_1'^2}{2m_1} + \frac{p_2'^2}{2m_2}. \tag{16}$$

The conservation of the total momentum is still Eqs. (3) and (4). It is then solved that

$$|p_1'| = \frac{|p_1|}{m_1 - m_2}[m_1 \cos\theta + g(\theta)] \tag{17}$$

and

$$|\bm{p}'_2| = \frac{2m_2|\bm{p}_1|\cos\varphi}{m_2 - m_1}. \tag{18}$$

The results are formally the same as Eqs. (12) and (13). Consequently, the two cases are the same as Eqs. (14) and (15). The only difference is that now the $m_1$ is of NKE and the $m_2$ is of PKE.

In Eqs. (12), (13), (17), and (18), there is a $m_2 - m_1$ in the denominators. Hence, the formulas are not applied to the case of $m_1 = m_2$. This case was discussed previously.[12]

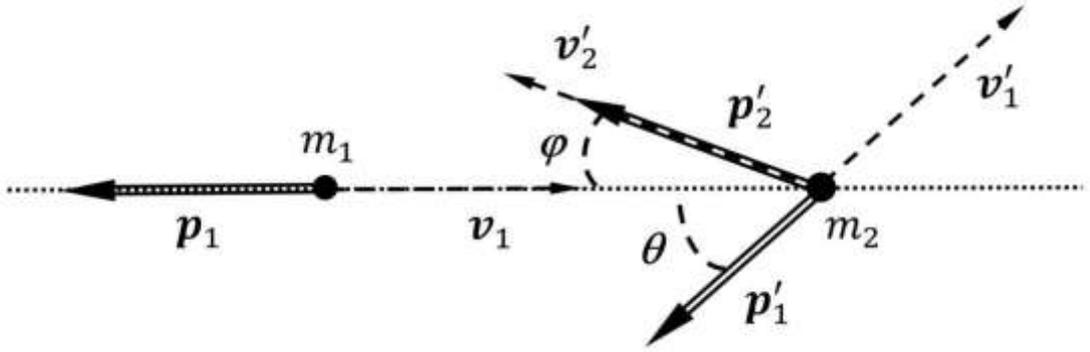

FIG. 3. An incident NKE particle with mass $m_1$ collides with a static PKE particle with $m_2$. Solid circles are the initial positions of the two particles. Dashed lines mean velocities. The double-solid lines are momenta. Please note that the directions of the velocity and momentum of a NKE particle are opposite to each other. When $m_1 = 0$, the incident particle is a dark photon.

**B. Relativistic motion**

The four-momentua of the two particles before the collision are respectively denoted as $p_{1i}$ and $p_{2i}$, and after the collision as $p_{1f}$ and $p_{2f}$. By the conservation of the total four-momenta, we can write

$$p_{1i} + p_{2i} - p_{1f} = p_{2f} \tag{19}$$

or

$$p_{1i} + p_{2i} - p_{2f} = p_{1f}. \tag{20}$$

The square of them leads to

$$m_1^2 + p_{1i} \cdot p_{2i} - p_{2i} \cdot p_{1f} - p_{1i} \cdot p_{1f} = 0 \tag{21}$$

or

$$m_2^2 + p_{1i} \cdot p_{2i} - p_{2i} \cdot p_{2f} - p_{1i} \cdot p_{2f} = 0. \tag{22}$$

B.1 Both $m_1$ and $m_2$ are of PKE

We first recall the collision between two PKE particles, as shown in Fig. 1. Before and after the collision, the incident particle's four-momentua are

$$p_{1i} = (E_1, \boldsymbol{p}_1), p_{1f} = (E_1', \boldsymbol{p}_1'), \tag{23}$$

and the target particle's four-momenta are

$$p_{2i} = (m_2, 0), p_{2f} = (E_2', \boldsymbol{p}_2'). \tag{24}$$

It is easily calculated that

$$p_{1i} \cdot p_{2i} = m_2 E_1, p_{2i} \cdot p_{1f} = m_2 E_1', p_{2i} \cdot p_{2f} = m_2 E_2', \tag{25}$$

$$p_{1i} \cdot p_{1f} = E_1 E_1' - |\boldsymbol{p}_1||\boldsymbol{p}_1'|\cos\theta, \tag{26}$$

and

$$p_{1i} \cdot p_{2f} = E_1 E_2' - |\boldsymbol{p}_1||\boldsymbol{p}_2'|\cos\varphi. \tag{27}$$

Substituting Eqs. (25)-(27) into Eq. (21), we obtain

$$E_1' = \frac{(E_1 + m_2)(m_1^2 + m_2 E_1) + g(\theta)\boldsymbol{p}_1^2 \cos\theta}{(E_1 + m_2)^2 - \boldsymbol{p}_1^2 \cos^2\theta}. \tag{28}$$

Substituting Eqs. (25)-(27) into Eq. (22), we get

$$E_2' = m_2 \frac{(E_1 + m_2)^2 + \boldsymbol{p}_1^2 \cos^2\varphi}{(E_1 + m_2)^2 - \boldsymbol{p}_1^2 \cos^2\varphi}. \tag{29}$$

If the low-momentum approximations are taken from these results, we will achieve the results in subsection A.1. In Eqs. (28) and (29), we take

$$E_1 \to m_1 + \boldsymbol{p}_1^2/2m_1, E_1' \to m_1 + \boldsymbol{p}_1'^2/2m_1, E_2' \to m_2 + \boldsymbol{p}_2'^2/2m_2.$$

Then, Eqs. (5) and (6) are retrieved. In Eq. (29), the range of the scattering angle of the target particle is $0 \leq \varphi < \pi$. By contrast, in the case of low-momentum motion, the corresponding scattering angle is within the range $0 \leq \varphi < \pi/2$, see Eq. (6). The reason is that at relativistic motion, the expansion of kinetic energy includes all even powers of momentum. This allows the outgoing particle to spread in the whole range of angle. At low-momentum motion, kinetic energy is only proportional to the square of momentum. Thus, under the condition of momentum conservation, there is a restriction for the scattering angle.

### B.2 $m_1$ is of PKE and $m_2$ is of NKE

The target particle is of NKE, see Fig. 2. Before and after the collision, the incident particle's four-momentua are

$$p_{1i} = (E_1, \boldsymbol{p}_1), \; p_{1f} = (E_1', \boldsymbol{p}_1'), \tag{30}$$

and the target particle's four-momenta are

$$p_{2i} = (-m_2, 0), \; p_{2f} = (-|E_2'|, \boldsymbol{p}_2'). \tag{31}$$

It is easily calculated that

$$p_{1i} \cdot p_{2i} = -m_2 E_1, \; p_{2i} \cdot p_{1f} = -m_2 E_1', \; p_{2i} \cdot p_{2f} = m_2 |E_2'|, \tag{32}$$

$$p_{1i} \cdot p_{1f} = E_1 E_1' - |\boldsymbol{p}_1||\boldsymbol{p}_1'|\cos\theta, \tag{33}$$

and

$$p_{1i} \cdot p_{2f} = -E_1|E_2'| - |\boldsymbol{p}_1||\boldsymbol{p}_2'|\cos\varphi. \tag{34}$$

Substituting Eqs. (32)-(34) into Eq. (21), we obtain

$$E_1' = \frac{(E_1 - m_2)(m_1^2 - m_2 E_1) + g(\theta)\boldsymbol{p}_1^2 \cos\theta}{(E_1 + m_2)^2 - \boldsymbol{p}_1^2 \cos^2\theta}. \tag{35}$$

Substituting Eqs. (32)-(34) into Eq. (22), we get

$$-|E_2'| = -|m_2| \frac{(E_1 + m_2)^2 + \boldsymbol{p}_1^2 \cos^2\varphi}{(E_1 + m_2)^2 - \boldsymbol{p}_1^2 \cos^2\varphi}. \tag{36}$$

Formally, one can let $m_2 \to -m_2$ in Eqs. (28) and (29) to get Eqs. (35) and (36).

For low-momentum approximation, we can take in Eqs. (35) and (36)

$$E_1 \to m_1 + \boldsymbol{p}_1^2/2m_1, \; E_1' \to m_1 + \boldsymbol{p}_1'^2/2m_1, \; -|E_2'| \to -m_2 - \boldsymbol{p}_2'^2/2m_2$$

so as to get Eqs. (12) and (13).

### B.3 $m_1$ is of NKE and $m_2$ is of PKE

The incident particle is of NKE and the target particle is of PKE, as shown in Fig. 3. Before the collision, the total momentum points to the left, and the direction remains after the collision.

Before and after the collision, the incident particle's four-momentua are

$$p_{1i} = (-|E_1|, \boldsymbol{p}_1), \; p_{1f} = (-|E_1'|, \boldsymbol{p}_1'), \tag{37}$$

and the target particle's four-momenta are

$$p_{2i} = (m_2, 0), \; p_{2f} = (E_2', \boldsymbol{p}_2') \tag{38}$$

It is easily calculated that

$$p_{1i} \cdot p_{2i} = -m_2 |E_1|, \quad p_{2i} \cdot p_{1f} = -m_2 |E_1'|, \quad p_{2i} \cdot p_{2f} = m_2 E_2', \tag{39}$$

$$p_{1i} \cdot p_{1f} = |E_1 E_1'| - |\mathbf{p}_1||\mathbf{p}_1'|\cos\theta, \tag{40}$$

and

$$p_{1i} \cdot p_{2f} = -|E_1 E_2'| - |\mathbf{p}_1||\mathbf{p}_2'|\cos\varphi. \tag{41}$$

Substituting Eqs. (39)-(41) into Eq. (21), we obtain

$$-|E_1'| = \frac{(|E_1|-m_2)(m_1^2 - m_2|E_1|) + g(\theta)\mathbf{p}_1^2 \cos\theta}{(|E_1|-m_2)^2 - \mathbf{p}_1^2 \cos^2\theta}. \tag{42}$$

Substituting Eqs. (39)-(41) into Eq. (22), we get

$$E_2' = m_2 \frac{(|E_1|-m_2)^2 + \mathbf{p}_1^2 \cos^2\varphi}{(|E_1|-m_2)^2 - \mathbf{p}_1^2 \cos^2\varphi}. \tag{43}$$

For low-momentum approximation, we can take in Eqs. (42) and (43)

$$-|E_1| \to -m_1 - \mathbf{p}_1^2/2m_1, \quad -|E_1'| \to -m_1 - \mathbf{p}_1'^2/2m_1, \quad E_2' \to m_2 + \mathbf{p}_2'^2/2m_2.$$

so as to get Eqs. (17) and (18).

The formulas describing the collision in Figs. 2 and 3 look the same form. This can be seen by comparing Eqs. (12) and (17), Eqs. (13) and (18), Eqs. (35) and (42), Eqs. (36) and (43). The reason is as follows: imagingin that the rest energy of both particles in one of the figures changes signs, it would becomes the other figure.

**C. The incident particle is a photon**

Let us consider the case that the incident particle is a photon, and the target particle can be of PKE or NKE. The collisions between two relativistic particles have been given in the previous subsection. We merely need to take the rest mass of the incident particle as 0 in the results of subsection B:

$$m_1 = 0. \tag{44}$$

Then, we have

$$E_1 = \omega, \quad E_1' = \omega', \quad \mathbf{p}_1^2 = \omega^2, \quad \mathbf{p}_1'^2 = \omega'^2. \tag{45}$$

Equation (7) is simplified to be

$$g(\theta) = \pm m_2 = \pm m. \tag{46}$$

C.1 The target particle is of PKE

This is the case of Compton scattering.[50,51] Substituting Eqs. (44)-(46) into Eq. (28), we obtain

$$\omega' = \frac{(\omega+m)m\omega \pm m\omega^2 \cos\theta}{(\omega+m)^2 - \omega^2 \cos^2\theta} = \frac{\omega}{1+(\omega/m)(1+\cos\theta)}. \tag{47}$$

After the photon is scattered, its frequency lowers. So, the minus sign in the numerator should be taken. This is the formula of Compton scattering. Substituting Eqs. (44)-(46) into Eq. (29), one gets

$$E_2' = m\frac{(\omega+m)^2 + \omega^2 \cos^2\varphi}{(\omega+m)^2 - \omega^2 \cos^2\varphi}. \tag{48}$$

Because the target particle is at rest before the collision, it gains energy after the collision.

C.2 The target particle is of NKE

We substitute Eqs. (44)-(46) into Eq. (28) and take $m \to -m$. The resut is

$$\omega' = -m\frac{(\omega-m)\omega \pm \omega^2 \cos\theta}{(\omega-m)^2 - \omega^2 \cos^2\theta} = \frac{\omega}{1-(\omega/m)(1-\cos\theta)}. \tag{49}$$

After the collision, the target particle gains a momentum, so that it obtains a NKE. Therefore, the photon's energy is increased. The frequency of the photon cannot be negative, so the plus sign should be taken in the numerator.

We substitute Eqs. (44)-(46) into Eq. (36) and take $m \to -m$ to get

$$-|E_2'| = -m\frac{(\omega-m)^2 + \omega^2 \cos^2\varphi}{(\omega-m)^2 - \omega^2 \cos^2\varphi}. \tag{50}$$

After the collision, the target particle gains a NKE.

**D. A possible way to increase the energy of a particle**

It is seen from Eq. (49) that if a photon is scattered by a NKE particle, its energy may have a great increase when the denominator is small.

For a PKE particle with a nonzero rest mass, subsection B.2 gives the formula for the increase of its energy after scattered by a NKE particle. We hope this energy increase be as high as possible. Let us discuss Eq. (36). The NKE increase of the target particle is the PKE increment of the incident particle, denoted as $\Delta K$.

$$\Delta K = |E_2'| - m_2 = \frac{2m_2 \boldsymbol{p}_1^2 \cos^2\varphi}{(E_1 - m_2)^2 - \boldsymbol{p}_1^2 \cos^2\varphi}. \tag{51}$$

We consider $\varphi \sim 0$. The quantity in the denominator is approximately $m_1^2 + m_2^2 - 2m_2 E_1$. As long as $E_1 < \frac{m_1^2 + m_2^2}{2m_2}$, $m_1$ will gain the increment $\Delta K$. From Eq. (51), the larger the $m_2$, the greater the $\Delta K$.

Figure 2 shows that a PKE particle is scattered by a NKE particle. We suggested in [1,12] a method to generate a free NKE electron, which was to use a photon or an electron to collide with a tunneling electron. The momentum of the outgoing NKE electron after the collision can be evaluated in terms of the detected momentum of the photon or the PKE particle.

Figures 2 and 3 can be combined. The outgoing NKE particle generated from Fig. 2 can be used to collide with a PKE particle, as shown in Fig. 3. In such a way, the energy of the PKE particle in Fig. 3 gets increased. This is called the first level of chain collisions. This outgoing PKE particle generated at the end of the first level can in turn collide with a NKE particles as shown in Fig. 2, and then, the outgoing NKE particle further collide with a PKE particle as shown in Fig. 3, which is the second level of chain collisions. At the end of the second level, the outgoing PKE particle gains an energy that is greater than that at the end of the first level. Let the chain collision continue. The more the level, the greater the energy the PKE particle can gain. This could be an inexpensive and clean way to boost particles' energies.

It is seen from Eq. (49) that if a photon collides with a NKE particle, then, when the exit angle of the NKE particle is nearly 0, its NKE increment could be very large. This particle with a very large NKE can be employed to collide with an massive particle. This way could be more efficient to increase the PKE of particles.

If a large number of NKE electrons can be produced by the way mentioned above, they can be introduced into an accelerator. After accelerated to high momenta, they can be employed to collide with PKE particles. On the one hand, the PKE particles would obtain high energy. On the other hand, observing the products of such collisions would be interesting.

The rest mass of an electron is much smaller than those of most other elementary particles. If the NKE particles with larger masses, say, protons, could be produced, the collisions described by Figs. 2 and 3 would increase PKE more efficiently. Up to now, generating NKE electrons by using tunneling electrons is the only way we can envisage.

## III. WHAT IS THE EQUILIBRIUM STATE WHEN A PKE AND A NKE IDEAL GASES MIX

We mentioned in the Introduction that there were three types of energy spectrum of systems. Now, consider that there is a PKE ideal gas and a NKE ideal gas. Their energy spectra are respectively type I and type II. So, they have positive and negative temperatures, respectively, when each of them is isolated. If they are mixed into each other and there is no interaction other than collisions, is it possible to achieve balance and what is the final state? We now study this problem.

The collisions between particles cause energy transfer, which was calculated in the previous section.

For convenience of discussion, we make some denotations. The distribution function of the PKE gas is denoted by $f$. For two PKE molecules, their momenta before a collision are $p_1$ and $p_2$, and they obey distribution functions $f_1(p_1)$ and $f_2(p_2)$, respectively. After the collision, their momenta change to be $p'_1$ and $p'_2$, and they obey distribution functions $f_1'(\boldsymbol{p}_1')$ and $f_2'(\boldsymbol{p}_2')$. The transition probability of the process $(\boldsymbol{p}_1, \boldsymbol{p}_2) \to (\boldsymbol{p}_1', \boldsymbol{p}_2')$ is $w(\boldsymbol{p}_1, \boldsymbol{p}_2; \boldsymbol{p}_1', \boldsymbol{p}_2')$. We have proved[52] that its reciprocal process $(\boldsymbol{p}_1', \boldsymbol{p}_2') \to (\boldsymbol{p}_1, \boldsymbol{p}_2)$ can also happen and the transition probability is the same: $w(\boldsymbol{p}_1, \boldsymbol{p}_2; \boldsymbol{p}_1', \boldsymbol{p}_2') = w(\boldsymbol{p}_1', \boldsymbol{p}_2'; \boldsymbol{p}_1, \boldsymbol{p}_2)$. The variation of the product of the distribution functions is $f_1(\boldsymbol{p}_1)f_2(\boldsymbol{p}_2) \to f_1'(\boldsymbol{p}_1')f_2'(\boldsymbol{p}_2')$. In equilibrium, the four functions $f_1, f_2, f_1', f_2'$ are the same, $f_1 = f_2 = f_1' = f_2' = f$, each of which is the distribution in equilibrium, and we have

$$f(\boldsymbol{p}_1)f(\boldsymbol{p}_2) = f(\boldsymbol{p}_1')f(\boldsymbol{p}_2').$$

This is the detailed balance.

The distribution function of the NKE gas is denoted by $g$. For two NKE molecules, their momenta before a collision are $\boldsymbol{q}_1$ and $\boldsymbol{q}_2$, and they obey distribution functions $g_1(\boldsymbol{q}_1)$ and $g_2(\boldsymbol{q}_2)$, respectively. After the collision, their momenta change to be $\boldsymbol{q}_1'$ and $\boldsymbol{q}_2'$, and they obey distribution functions $g_1'(\boldsymbol{q}_1')$ and $g_2'(\boldsymbol{q}_2')$. The transition probability of the process $(\boldsymbol{q}_1, \boldsymbol{q}_2) \to (\boldsymbol{q}_1', \boldsymbol{q}_2')$ is $w(\boldsymbol{q}_1, \boldsymbol{q}_2; \boldsymbol{q}_1', \boldsymbol{q}_2')$. Obviously, like the case of a PKE ideal gas, the reciprocal process $(\boldsymbol{q}_1', \boldsymbol{q}_2') \to (\boldsymbol{q}_1, \boldsymbol{q}_2)$ can also occur and the transition probability is the same: $w(\boldsymbol{q}_1, \boldsymbol{q}_2; \boldsymbol{q}_1', \boldsymbol{q}_2') = w(\boldsymbol{q}_1', \boldsymbol{q}_2'; \boldsymbol{q}_1, \boldsymbol{q}_2)$. The variation of the product of the distribution functions is $g_1(\boldsymbol{q}_1)g_2(\boldsymbol{q}_2) \to g_1'(\boldsymbol{q}_1')g_2'(\boldsymbol{q}_2')$. In equilibrium, the four functions $g_1, g_2, g_1', g_2'$ are the same, $g_1 = g_2 = g_1' = g_2' = g$, each of which is the distribution in equilibrium, and there should also be the detailed balance,

$$g(\boldsymbol{q}_1)g(\boldsymbol{q}_2) = g(\boldsymbol{q}_1')g(\boldsymbol{q}_2').$$

Suppose that a PKE ideal gas A is in its equilibrium state with a temperature $T_A > 0$, and a NKE ideal gas B is in its equilibrium state with a temperature $T_B < 0$. We mix them. It is natural that a collision between a pair of PKE molecules does not violate the equilibrium state of A, and that a collision between a pair of NKE molecules does not violate the equilibrium state of B. A collision between a PKE and a NKE molecules makes their momenta change, $(\boldsymbol{p}_1, \boldsymbol{q}_2) \to (\boldsymbol{p}_1', \boldsymbol{q}_2')$, which was studied in the previous section. We have assumed that before the mix, the two ideal gases have been in their respective equilibrium states. That means that in the beginning, the two molecules before a collision are in their own equilibrium distribution $f_1 = f, g_2 = g$. After the collsion, $f(\boldsymbol{p}_1)g(\boldsymbol{q}_2) \to f_1'(\boldsymbol{p}_1')g_2'(\boldsymbol{q}_2')$. There could be two cases for the $f_1'(\boldsymbol{p}_1')$ and $g_2'(\boldsymbol{q}_2')$.

One is that they are still equilibrium distributions, $f_1' = f, g_2' = g$, and $f(\boldsymbol{p}_1)g(\boldsymbol{q}_2) = f_1(\boldsymbol{p}_1')g_2(\boldsymbol{q}_2')$. The collisions between a PKE and a NKE molecules does not violate the distributions of the PKE and NKE ideal gases. In the mixed system, the PKE gas keeps its equilibrium with temperature $T_A$ and the NKE gas keeps its equilibrium with temperature $T_B$.

The other cases is that $f_1'(\boldsymbol{p}_1')$ and $g_2'(\boldsymbol{q}_2')$ are not the equilibrium distributions. We consider subsequent collisions. When the $\boldsymbol{p}_1'$ molecule collides with another PKE molecule, the PKE gas will tend to be at equilibrium. When the $\boldsymbol{q}_2'$ molecule collides with another NKE molecule, the NKE gas will tend to be equilibrium. Moreover, the reciprocal process $(\boldsymbol{p}_1', \boldsymbol{q}_2') \to (\boldsymbol{p}_1, \boldsymbol{q}_2)$ would make the gases go back to equilibrium.

The conclusion is that after the mix, the PKE gas keeps its equilibrium state with temperature $T_A > 0$ and the NKE gas keeps its equilibrium state with temperature $T_B < 0$.

Now, if an isolated PKE ideal gas A is in a nonequilibrium state. With time going, it will approach an equilibrium state with a temperature $T_A > 0$. If an isolated NKE ideal gas B is in a nonequilibrium state. With time going, it will approach an equilibrium state with a temperature $T_B < 0$. Suppose that the two ideal gases both in nonequilibrium states are mixed. What will be the final state? The collisions between the PKE molecules will make the PKE part tend to be the equilibrium state with temperature $T_A > 0$. The collisions between the NKE molecules will make the NKE part tend to be the equilibrium state with temperature $T_B < 0$. The analysis of the collisions between a PKE and a NKE molecules is just the same as mentioned above.

The conclusion is that evebtually, the PKE part will be in the equilibrium state with $T_A > 0$ and the NKE part will be in the equilibrium state with temperature $T_B < 0$.

The analysis in this section ought to be verified. We suggest that numerical simulation be carried out to confirm the conclusion in this section.

The analysis in this section should be valid for all the cases when a type-I ideal gas and a type-II ideal gas are mixed.

## IV. COSMIC DARK RADIATION BACKGROUND AND GRAVITY POTENTIAL BACKGROUND

### A. Cosmic Dark Radiation Background

It is well-known that there is a Cosmological Microwave Background (CMB) in the universe. The CMB is of a temperature about 2.73K.[53,54,55] This background is isotropic and uniform everywhere. The CMB is believed the relic of the Big Band.

It is now generally believed that there is dark energy in the universe. In our previous work,[11] we proposed that dark energy was dark radiation, which was composed of dark photons, that is, negative energy photons. Then, like photons, there should be a background of dark radiation in the universe, called the Cosmic Dark Radiation Background (CDRB). The CMB and CDRB are in the same universe. It is likely that collisions between photons and dark photons take place. The CMB has a type-I energy spectrum, while the CDRB has a type-II energy spectrum. In the previous section, we have analyzed that when a type-I ideal gas and a type-II ideal gas are mixed, each part will reach its own equilibrium state with a temperature, just as it is isolated. Consequently, although the CMB and CDRB are mixed in the universe, each reaches its own equilibrium state. The equilibrium temperature of the CDRB is a negative number. This CDRB is believed to contribute a part to the dark energy in the universe.

### B. Gravity Potential Background

There is gravitational force between objects. Every celestial body in the universe is subjected to the gravitational pull of all other celestial bodies. For example, the Earth, which is attracted by the Sun and revolves around the Sun, is also affected by the gravitational pull of its neighboring celestial bodies such as the Moon, Venus, Mercury, Mars, etc., and is also affected by the gravitational action of Proxima Centauri, the closest star to the solar system, and the gravitational action of all other celestial bodies in the entire Milky Way, as well as the gravitational action of celestial bodies in exogalaxies.

In short, the Earth has a gravitational potential energy. In this article, when mentioning potential energy, we always set the zero of potential energy at infinity.

Of course, celestial bodies outside the solar system are very far from the Earth, and their gravitational pull on Earth is very weak. However, considering that the universe is large enough, the sum of the gravitational potential energy of all other celestial bodies in the universe except the solar system for the Earth should be a nonzero value. The gravitational potential decays with distance to the first power. Note that this differs from the discussion of Olbers' paradox, since the luminosity of a star decays with distance squared.[56] If the universe were infinite, the total gravitational potential of the Earth would be infinite. But modern cosmological models suggest that the universe is not infinite, so the total gravitational potential of the Earth by the universe is a finite value.

Similarly, other celestial bodies, such as the Sun, are also subjected to the gravitational pull of all other objects in the universe besides the solar system. This total gravitational potential should be a finite value that is not zero.

The motion of a celestial body is mainly governed by the gravitational action of neighboring objects. For example, the movement of the Earth is mainly affected by the Sun, Moon, and planets within the solar system. The sum of the gravitational potential of other celestial bodies on Earth is called the gravity potential background (GPB).

According to the cosmology principles,[56] matter in the universe is uniformly distributed and isotropic on large scales. Thus, the GPB should be uniform and equal everywhere in the universe. Different celestial bodies are subjected to this gravitational background equally. Consequently, every celestial body cannot perceive itself in such a background. For one celestial body, only the gravitational action of its neighboring celestial bodies will have a non-negligible effect and govern its movement.

The gravitational potential energy is negative. The universe has a negative gravitational potential energy. The sum of GPB of the whole universe is a finite gravitational potential energy.

As the universe expands, the distance between celestial bodies in the universe is slowly increased. The GPB will get smaller, keeping uniform everywhere.

It is thought that there are two origins for the vacuum zero-point energy of massive particles: harmonic oscillator models of low-momentum particles and second quantization of relativistic free particles, see Table I in [8]. We have argued that the zero-point energy of the former does not exist. In future work, we will demonstrate that zero-point energy will not appear after correct second quantization of relativistic free particles. In another work,[11] we showed that the total zero-point energy of radiation and dark radiation in the full-space was zero. The zero-point energy is thought to be related to the cosmological constant in the field equations of general relativity. Our work shows that there is no so-called vacuum zero energy in the whole universe. We suspect that the cosmological constants, if it exists, may be related to the gravitational potential background and/or the dark radiation background.

**C. Negative energy is predominating in the universe**

From the CDRB and GPB we realize that our universe is full of negative energy.

Celestial bodies form stable systems in terms of gravity between each other, such as the Solar System, Milky Way, Local Group of Galaxies, Virgo Supercluster, and so on. For each of the stable systems, the total energy is negative.

If a stable system is composed of PKE substances, it is called a PKE system. If a stable system is composed of NKE substances, it is called a NKE system.

In our observable universe, almost all of the PKE substances compose stable systems in motion. From macroscopic galaxies and galaxy clusters to microscopic atomic nuclei. The total energy of an interacting system is kinetic energy plus potential energy. For PKE systems, kinetic energy is always positive, while potential energies can be positive or negative. The condition for a PKE system to be stable is that the total attraction potential energy exceeds the total PKE, and the total energy of every stable PKE systems must be negative. This again reflects that negative energy is predominating in the universe.

All the substances we can observe are PKE ones, and they emit and absorb photons. People are used to this fact, but not used to that the kinetic energy and radiation can also be negative.[11]

The total energy of the universe is negative, and the universe should evolve from a non-equilibrium state to an equilibrium state on a large scale. A stable PKE system must have negative total energy. This prompts us that the origin of the universe might have negative energy. The stable PKE system with negative total energy observed now should originate from negative energy.

In the mext section, we propose a possible mechanism that NKE substances metamorphose to be PKE ones.

## V. A POSSIBLE MECHANICS THAT STABLE PKE SYSTEMS FORM

Suppose that there are two NKE bodies A and B far away from each other with gravity between them. Let A be static at the origin. The body B with a mass $m$ is moving, subject to a force $f = -\dfrac{k}{r^3} r$. Its equation of motion is $f = \dfrac{d\bm{p}}{dt}$. Because gravity is a central force, the mechanical energy of B is conserved in its motion. Let us consider the following process.

Initially, because B is of NKE, it has a relation $\bm{p} = -m\bm{v} = -m\dfrac{d\bm{r}}{dt}$,[4] so that $f = -m\dfrac{d^2\bm{r}}{dt^2}$. We have $m\dfrac{d^2 r}{dt^2} = m\dfrac{dv}{dt} = \dfrac{k}{r^2}$. This equation is recast to be $\dfrac{m}{2}\dfrac{dv^2}{dt} = k\dfrac{d}{dt}\dfrac{1}{r}$. As it is closer to the origin, its gravity potential gets stronger, and correspondingly, its NKE becomes weaker. There is a distance $r = r_0$ at which the NKE of B becomes zero: $\dfrac{m}{2}\int_v^0 dv'^2 = k\int_r^{r_0} d\dfrac{1}{r'}$. By integration, we get

$$r > r_0, \quad -\frac{1}{2}mv^2 = \frac{k}{r} - \frac{k}{r_0}. \tag{52}$$

As $r > r_0$, both potential and kinetic energy are negative.

At the point $r = r_0$, the velocity of B becomes zero. Because it is still subject to an attractive force, it continues moving toward the origin. As $r < r_0$, the gravity potential further becomes stronger, such that the kinetic energy have to be positive. For a PKE body, the equation of motion becomes $f = m\frac{d^2 r}{dt^2}$, that is to say, $m\frac{d^2 r}{dt^2} = m\frac{dv}{dt} = -\frac{k}{r^2}$.

This equation is recast to be $\frac{m}{2}\frac{dv^2}{dt} = -k\frac{d}{dt}\frac{1}{r}$. It follows that $\frac{m}{2}\int_0^v dv'^2 = -k\int_{r_0}^r d\frac{1}{r'}$. By integration, we obtain

$$r < r_0, \quad \frac{1}{2}mv^2 = \frac{k}{r} - \frac{k}{r_0}. \tag{53}$$

As $r < r_0$, the potential is still negative but the kinetic energy is positive. The total mechanical energy $-\frac{k}{r_0}$ remains unchanged.

We are familiar with the cases that two PKE bodies consititute a stable system in terms of gravity between them, but the toal mechanical energy of such a system is negative.

Thus, we have proposed a mechanism that two NKE bodies with gravity between them can constitute a stable PKE system.

Now, consider that both bodies are moving. Their masses, position vectors, and momenta are respectively denoted as $m_1$ and $m_2$, $r_1$ and $r_2$, $p_1$ and $p_2$. Their total mass $M$ and reduced mass $\mu$ are defined by

$$M = m_1 + m_2, \quad \frac{1}{\mu} = \frac{1}{m_1} + \frac{1}{m_2}. \tag{54}$$

Their total momentum and relative momentum are defined by

$$\boldsymbol{P} = \boldsymbol{p}_1 + \boldsymbol{p}_2, \quad \boldsymbol{p} = \boldsymbol{p}_1 - \boldsymbol{p}_2. \tag{55}$$

When they are sufficiently far away, their total NKE is

$$E_{(-)\text{T}} = -\frac{p_1^2}{2m_1} - \frac{p_2^2}{2m_2} = -\frac{P^2}{2M} - \frac{p^2}{2\mu}. \tag{56}$$

When they get closer, it is possible for them to combine into a stable PKE system by gravity between them. In the course of combination, the centroid kinetic energy remains unchanged. The total energy of the stable system is

$$E_{(-)T} = -\frac{P^2}{2M} + E, \tag{57}$$

where $E$ is the total mechanical energy inside the system and is the sum of kinetic energy $T$ and potential $V$.

$$E = T + V < 0. \tag{58a}$$

Here, $T > 0$ and $V < 0$. The total mechanical energy is negative. Comparison of Eqs. (56) and (57) leads to

$$E = -\frac{p^2}{2\mu}. \tag{58b}$$

In the course of combination, there is no energy exchange between the system and enviroment.

We turn to consider two PKE bodies. When they are far from each other, the gravity between them can be neglected and their total energy is

$$E_{(+)T} = \frac{p_1^2}{2m_1} + \frac{p_2^2}{2m_2} = \frac{P^2}{2M} + \frac{p^2}{2\mu}. \tag{59}$$

Suppose that they get closer. When gravity cannot be neglected and they constitute a stable system by means of gravity, the total energy of the system should be

$$E_{(+)} = \frac{P^2}{2M} + E. \tag{60}$$

The centroid kinetic energy must remain unchanged. The $E$ in Eq. (60) should be the total mechanical energy of the stable system so that it must be negative. Obviously, it is impossible for Eqs. (60) and (59) to be equal. That is to say, two PKE bodies cannot constitute a stable system by means of gravity between them, if they do not exchange energy with environment.

If at the very beginning, the universe is full of PKE objects, it would be almost impossible for all of them to form stable systems as seen today. If two PKE objects form a stable PKE system by gravity between them, the collision with the third object is necessary so as to transfer part of the PKE to the third object. As a result, the PKE of this third object increases. That is to say, there are always some PKE objects that cannot constitute a stable system with other objects by gravity.

In short, in the macroscopic field, all NKE objects can constitute stable PKE systems by gravity. In contrast, if all macroscopic objects are of PKE, at least some of them cannot participate in constituting stable systems by gravity.

We turn to microscopic particles.

Let us take a proton and an electron as an example. Because they carry contrary electric charges, they can attract each other. A PKE proton and a PKE electon constitute a stable hydrogen atom by means of Coulomb attraction between them, which is the case we are familiar with. A hygrogen atom is a PKE system, because inside the hydrogen, both the proton and electron have PKE.

The discrepancy of a macroscopic system and a microscopic system is that in the latter, there are a series of stationary states with discrete energy levels.

Now, we imagine that there are a NKE proton and a NKE electron. When they are far away from each other, both are free particles. Microscopic particles are of wave-particle duality. There is a nonzero probability for a microscopic particle to arrive at any possible state, due to its wave nature. If the NKE proton and NKE electron are close enough, there will be some probability for them to transit to an enengy level of a hydrogen atom. This hydrogen atom is a stable PKE system.

Let us evaluate the probability that the two free NKE particles transit to an energy level of the hydrogen atom. The mass, position vector, and momentum of the proton are denoted by $m_1$, $r_1$, and $p_1$, and those of the electron denoted by $m_2$, $r_2$, and $p_2$. Their total mass $M$ and reduced mass $\mu$ are defined by Eq. (54).

The initial state is that the two particles are far away from each other so that the interaction between them can be neglected. The Hamiltonian and wave function of the initial state are marked by a subscript i. The Hamiltonian of the initial state is

$$H_{(-)i} = \frac{\hbar^2}{2m_1}\nabla_1^2 + \frac{\hbar^2}{2m_2}\nabla_2^2 = \frac{\hbar^2}{2M}\nabla_R^2 + \frac{\hbar^2}{2\mu}\nabla_r^2. \tag{61}$$

The wave functions of the two free particles are respectively

$$\varphi_1(r_1) = \frac{1}{(2\pi)^{3/2}} e^{-ip_1 \cdot r_1/\hbar} \tag{62a}$$

and

$$\varphi_2(r_2) = \frac{1}{(2\pi)^{3/2}} e^{-ip_2 \cdot r_2/\hbar}. \tag{62b}$$

When using the total mass and reduced mass, we regard the total mass and reduced mass as two free particles and their wave functions are respectively

$$\psi_1(R) = \frac{1}{(2\pi)^{3/2}} e^{-iK \cdot R/\hbar} \tag{63a}$$

and

$$\psi_2(r) = \frac{1}{(2\pi)^{3/2}} e^{-ik \cdot r/\hbar}. \tag{63b}$$

The stationary wave function of the initial state is

$$\psi_i = \frac{1}{(2\pi)^3} e^{-i(p_1 \cdot r_1 + p_2 \cdot r_2)} = \frac{1}{(2\pi)^3} e^{-i(K \cdot R + k \cdot r)}. \tag{64}$$

The total energy of the system is

$$E_{(-)i} = -\frac{\hbar^2 k_1^2}{2m_1} - \frac{\hbar^2 k_2^2}{2m_2} = -\frac{\hbar^2}{2M}K^2 - \frac{\hbar^2}{2\mu}k^2. \tag{65}$$

The final state is that the two particles, by means of Coulomb attraction between them, constitute a hydrogen atom in which both the proton and electron are of PKE. The Hamiltonian and wave function of the final state are marked by a subscript f. Please note that the transition does not change the centroid momentum as well as centroid NKE if the system does not suffer interaction from outside. The Hamiltonian of the final state is

$$H_{(-)f} = \frac{\hbar^2}{2M}\nabla_R^2 - \frac{\hbar^2}{2\mu}\nabla_r^2 - \frac{e^2}{4\pi\varepsilon_0 r^2}. \tag{66}$$

This Hamiltonian is composed of two parts. One is the centroid NKE.

$$H_{(-)f1} = \frac{\hbar^2}{2M}\nabla_R^2. \tag{67}$$

Its wave function is still (63a). The other part is

$$H_{f2} = -\frac{\hbar^2}{2\mu}\nabla_r^2 - \frac{e^2}{4\pi\varepsilon_0 r^2}. \tag{67b}$$

This is the motion of the reduced mass $\mu$ subject to the Coulomb attraction. Its wave function is that of a hydrogen atom with mass $\mu$.

$$\psi_{nlm}(\mathbf{r}) = R_n(r)P_l^m(\cos\theta)e^{im\varphi}. \tag{68}$$

The stationary wave function of the final state of the system is

$$\psi_f = \frac{1}{(2\pi)^{3/2}}e^{-i\mathbf{K}\cdot\mathbf{R}}\psi_{nlm}(\mathbf{r}). \tag{69}$$

The total energy is the centroid NKE plus the energy of the hydrogen atom.

$$E_{(-)f} = E_{(-)K} + E_{nlm} = -\frac{\hbar^2 K^2}{2M} - \frac{e^2}{8\pi\varepsilon_0 an^2}, a = \frac{4\pi\varepsilon_0\hbar^2}{\mu e^2}. \tag{70}$$

We use an expansion formula of plane wave,

$$e^{i\mathbf{k}\cdot\mathbf{r}} = \sum_{l=0}^{\infty}i^l(2l+1)j_l(kr)P_l(\cos\theta), \tag{71}$$

where $j$ means spherical Bessel function, and $\theta$ is the angle between the $\mathbf{k}$ and $\mathbf{r}$. Then, we are able to evaluate the transition amplitude from the initial to final states.

$$\langle\psi_f|\psi_i\rangle = \frac{i^l\delta_{m0}}{(2\pi)^{1/2}}\int_0^\infty r^2 dr j_l(kr)R_n(r). \tag{72}$$

For instance, the ground state wave function is

$$\psi_{100}(r) = \frac{1}{\sqrt{\pi a^3}}e^{-r/a} = R_1(r). \tag{73}$$

So, the transition amplitude from the initial state to the ground state of the hydrogen atom is

$$\langle \psi_f | \psi_i \rangle = \frac{1}{\sqrt{2a\pi}(k^2 a^2 + 1)^2}. \tag{74}$$

There is also a probability for the particle to transit from the initial state to an excited state of the hydrogen atom. If so, spontaneous transition will occur, such that the hydrogen atom will transit to the ground state or another state with lower energy level, emitting photons.

The total energy should be conserved. Comparing Eqs. (65) and (70), we know that

$$-\frac{k^2}{2\mu} = -\frac{e^2}{8\pi\varepsilon_0 a n^2}. \tag{75}$$

It is noted that the energy of a stable hydrogen atom ranges from zero to $E_1 = -\frac{e^2}{8\pi\varepsilon_0 a}$, the latter being the ground state energy. This provides a restriction to the energy of the initial state:

$$E_1 = -\frac{e^2}{8\pi\varepsilon_0 a} < -\frac{\hbar^2 k^2}{2\mu} = -\frac{e^2}{8\pi\varepsilon_0 a n^2} < 0, \tag{76}$$

if there is no energy exchange with environment during the transition described by Eq. (72).

The above discussion is for low-momentum motion. Strictly, relativistic motion ought to be taken into account. The energies of two free particles are respectively $\varepsilon_1 = -\sqrt{m_1^2 + p_1^2}$ and $\varepsilon_2 = -\sqrt{m_2^2 + p_2^2}$. The transition to a stable hydrogen atom is not easily written. This is because the relativistic motion of the two particles cannot be equivalent to that of a total mass and a reduced mass. Relativistic momentum is not linearly proportional to velocity. So far, there has not been a two-particle Dirac equation.[57] Therefore, we are unable to make quantitative analysis as in the case of low-momentum motion. It is estimated that the binding energy of a hydrogen atom is about 10eV, and the rest mass of an electron is about 0.5MeV. They are negligible compared to the rest mass of a proton, about 938MeV. The negative rest energy of a proton metamorphoses to be positive rest energy and then the proton participates in constituting a PKE hydrogen atom. In this course, some negative energy would be released to environment.

Let us consider the case that a PKE proton and a PKE electron constitute a stable hydrogen atom. The initial state is that they are far away from each other, and the Hamiltonian is

$$H_{(+)i} = -\frac{\hbar^2}{2m_1}\nabla_1^2 - \frac{\hbar^2}{2m_2}\nabla_2^2 = -\frac{\hbar^2}{2M}\nabla_R^2 - \frac{\hbar^2}{2\mu}\nabla_r^2. \tag{77}$$

The initial wave function is again Eqs. (62)-(64). the total energy is

$$E_{(+)i} = \frac{\hbar^2 k_1^2}{2m_1} + \frac{\hbar^2 k_2^2}{2m_2} = \frac{\hbar^2}{2M}K^2 + \frac{\hbar^2}{2\mu}k^2. \tag{78}$$

When they become closer, they can combine into a stable hydrogen atom in terms of Coulomb attraction between them. The Hamiltonian of the final state is

$$H_{(+)f} = -\frac{\hbar^2}{2M}\nabla_R^2 - \frac{\hbar^2}{2\mu}\nabla_r^2 - \frac{e^2}{4\pi\varepsilon_0 r^2}. \tag{79}$$

If the centroid energy is not changed, the first term in Eq. (79) is the same as the centroid Hamiltonian in Eq. (78), which has an inverse sign of Eq. (67a). The wave function of the final state is again Eq. (69). The total energy of the system is the energy of the centroid plus that of a hydrogen atom.

$$E_{(+)f} = E_{(+)K} + E_{nlm} = \frac{\hbar^2 K^2}{2M} - \frac{e^2}{8\pi\varepsilon_0 a n^2}. \tag{80}$$

The transition amplitude from the initial state to final state is again evaluated by Eq. (72). We compare Eqs. (77) and (79). The term $-\frac{\hbar^2}{2\mu}\nabla_r^2$ in Eq. (77) is a positive energy, while the $-\frac{\hbar^2}{2\mu}\nabla_r^2 - \frac{e^2}{4\pi\varepsilon_0 r^2}$ in Eq. (79) is a negative energy. Therefore, if the transition from the initial state Eq. (77) to the final state Eq. (79) occurs, the system has to emit photons.

In summary, in microscopic field, NKE particles can constitute a stable PKE system in terms of attractions between them, with the total energy unchanged or accompanied by releasing negative energies. In contrast, if PKE particles constitute a PKE system in terms of attractions between them, there must be photons released.

One point should be stressed. After NKE particles constitute a stable PKE system in terms of attractions between them, the centroid is still of NKE. Such a system is called a PKE system with a centroid NKE. Two PKE systems with centroid NKE can, by means of attractions between them, further constitute a larger PKE system with centroid NKE. For example, two hydrogen atoms with centroid NKEs can constitute a stable hydrogen molecule. For this molecule, each hydrogen is of PKE, and the centroid of the molecule is of NKE.

Similarly, smaller molecules can constitute a larger molecule in terms of attractive interactions between them, and so on. In general, smaller PKE systems with centroid NKE can constitute a larger stable PKE system with centroid NKE by means of attraction interactions between them. In the light of this mechanism, larger and larger PKE systems with centroid NKE can be formed, gradually from microscopic to macroscopic levels. In each level, the final product has a centroid NKE, while its interior is totally PKE. That is to say, inside the final product, one can only see the PKE behaviors of the system, being no aware of that the centroid is of NKE. For example, if at a level the final product is the Earth, the matter inside the Earth all behaves with PKE, but the mass center of the Earth is of NKE. However, the Earth is now not an isolated body, but participates in the solar system with the Sun and other planets. Inside the solar system, everything is of PKE behaviors. Hence, the Earth, as a component of the solar system, has a PKE mass center. Only the solar system as a whole, if it is isolated, could have a centroid NKE. It is well-known that the solar system is a component of the Milky Way, so that the mass center of the solar system has a PKE. The Milky Way is a component of the Local Group. The Local Group is a component of the Supercluster.

We review the meaning of the NKE.

According to Newtonian mechanics, if a body with mass $m$ has a velocity $v$, it will have a momentum

$$\boldsymbol{p} = m\boldsymbol{v} . \tag{81}$$

The directions of momentum and velocity are the same. We have pointed out that Eq. (81) is the relationship for PKE bodies.[4] For a NKE object, the relationship between momentum and velocity is

$$\boldsymbol{p} = -m\boldsymbol{v} . \tag{82}$$

The directions of momentum and velocity are opposite to each other.

Therefore, the way to make clear whether a material system has a PKE or a NKE is to observe whether the directions of its momentum and velocity are the same or opposite.

Velocity is a kinematic quantity, while momentum is a dynamic quantity. For a macroscopic object, the way of measuring its velocity is different from that of measuring its momentum. One can measure the velocity of an object without affecting its motion, e.g., measuring the velocity of the Earth. When we are talking about a body's velocity, we refers to its velocity relative to its neighboring objects or to one reference point. For instance, the speed of the Earth's refers to its motion relative to the Sun. When it comes to the speed of the solar system, it refers to its motion relative to the center of the Milky Way. However, measuring the momentum of an object relies on its interaction with other objects, such as collision with another object.

If an object definitely has a PKE and both its velocity and momentum can be measured, its momentum-velocity relation is Eq. (81). If we are able to measure an object's velocity but not its momentum, then, the momentum could be determined by either Eq. (81) or Eq. (82). Without acting on it by another object, one is unable to determine its momentum's direction.

If the Earth is isolated, we are unable to determine its momentum's direction, because it is impossible for us to utilize another body to collide with the Earth. The same discussion applies to the Milky Way if its is isolated.

Fortunately, as having been analized above, we can safely say that for the celestial systems such as the Earth, Solar System, Milky Way, Local Group of Galaxies, and Virgo Supercluster, each of them has a centroid PKE, because each is a component of a larger stable system that is formed by means of gravity. Only the highest-level system, if it is isolated as a whole, could have a centroid NKE.

# VI. A QUALITATIVE EXPLANATION TO THE PROPORTIONS OF THE FOUR INGREDIENTS IN THE UNIVERSE

Let us first recalled what the standard model of cosmology says about the early time of the universe. "The current standard model of cosmology posits that at a very early time, the universe was nearly homogeneous and isotropic, radiation-dominated, and nearly flat."[58] Another statement is that "it should be a good approximation to treat the universe during this era as if it were filled purely with radiation, with essentially no matter at all. …… The age of pure radiation actually began only at the end of the first few minutes, when the temperature had dropped below a few thousand million degrees Kelvin. At earlier times matter *was* important, but matter of a kind very different from that of which our present universe is composed. However, before we look that far back, let us first consider briefly the true era of radiation, from the end of the first few minutes up to the time, a few hundred thousand years later, when matter again became more important than radiation. …… Consider a time in the past when the temperature was above ten million million degrees $10^{13}$ K, the threshold temperature for neutrons and protons. At that time the universe would have contained plenty of nuclear particles and antiparticles, about as many as photons."[59] In short, the universe was initially filled with radiation; Matter is produced from radiation, mainly neutrons and protons; The amount of radiation and that of produced matter have about the same magnitude; The time period is from the end of a few minutes to hundreds of thousands of years. We compare this information with Eq. (1).

Equation (1) tell us that the ratio of radiation to matter now is about $\Omega_{R0}/\Omega_{M0} \sim 10^{-3}$. The first feeling is that when particles combine into a stable system, some photons may be emitted, which is quantitatively much less than the particles themselves. For example, when a neutron and a proton form a stable nucleus, the nucleus emits γ rays with an energy of 2.231 MeV, while a neutron has a mass of about 940 MeV. In this way, the ratio of the energy released to the amount of matter is about $10^{-3}$. This seems reasonable. However, this is questionable. The premise that almost all of the radiation in the early universe was converted into matter may not be true. There are two reasons for this suspicion. One reason is that there is a part of radiation that could not be converted into matter. We believe that there was a certain probability that photons could collide each other, and among all the collisions, there was a certain probability that photons could annihilate to produce particle-antiparticle pairs. Although we do not know what was the probabilities, they wear certainly not nearly 100%. It was not possible for all photons to collide into particles. There must have been some photons that cannot be converted into particles. The second reason is that while a pair of photons collided to produce a particle-antiparticle pair, the reverse process also existed, that is, the collision of a pair of particle-antiparticle produced a pair of photons. It was possible for the process and its reverse process to reach a certain thermal equilibrium. This means that there was always a part of photons that could not be converted into matter. This part of energy should be about the same order of magnitude as the amount of matter. This conclusion is inconsistent with $\Omega_{R0}/\Omega_{M0} \sim 10^{-3}$ in Eq. (1).

We now speculate about the evolution of the early universe as follows.

At the earliest moment of the universe, there was only a high density of dark energy, that is, it was filled with dark photons. The universe had a very high negative temperature. Negative energy photons collided with each other frequently. A pair of negative photons collided and annihilated, producing a pair of NKE particle-antiparticle,[11] which is just like that a pair of photons collide to produce a PKE particle-antiparticle pair. The NKE matter is dark matter. Therefore, dark matter was produced from dark energy. There was a certain probability that negative photon collided to produce NKE particles. At the same time, there would be a pair of NKE particle-antiparticle colliding and annihilating into a pair of negative photons. Therefore, merely a small part of dark radiation could be converted into NKE matter. In general, the amount of dark matter produced should be less than the amount of dark energy. As the universe expands, the density of dark radiation decreases. The probability of negative photon collisions producing NKE particles basically approached to zero. Therefore, when the negative temperature in the universe "cooled" to a certain value, this transition between dark photons and NKE particles stopped, and the ratio of dark energy and dark matter became fixed. There should be $\Omega_{\Lambda 0} > \Omega_{DM0}$. The CDRB proposed in Section IV might be the relic of the dark radiation in the early universe.

Once the NKE particles were generated, the mechanism proposed in Section V could play a role, i.e., they could combine into stable systems by means of interactions between them. There was certainly some probability for two or more NKE particles to be close enough to combine. So, only a fraction of all the NKE particles could be able to compose stable systems. In constituting stable systems, there could be two ways. One way is that NKE particles could combine into stable PKE systems by means of attractions between them, a mechanism mentioned in the last section. For example, a proton and a neutron made up a deuterium nucleus, and a proton and an electron made up a stable hydrogen atom. In this way, matter was produced. That is to say, matter was produced from dark matter. Another way is that NKE particles combined into NKE systems by means of repulsive interactions between them. For example, two NKE electrons could form a stable NKE electron pairs by Coulomb repulsion between them,[1] and similarly, two NKE protons could form a stable NKE proton pairs. A NKE electron and a NKE proton could form a dark hydrogen atom, see Table 2 in [4]. The stable systems produced in this way were still dark matter. In both ways, some dark photons might be released in constituting stable systems. It is easily understood that merely a fairly small proportion of dark matter could be converted into matter. As the universe expands, the density of NKE decreases, and the probability that NKE particles composed stable systems approached zero. Consequently, the ratio of dark matter to matter became fixed. Based on the analysis, we think that there should be $\Omega_{DM0} > \Omega_{M0}$.

The energy levels of microsystems are discrete. When two free NKE particles transited to a state of a stable PKE system, they did not necessarily transit to the ground state of the system, and there was a certain probability of transitioning to any of the energy levels. When the final state was an excited state, spontaneous transition would occur, from the excited state to the ground state or another lower energy level, emitting photons. Thus, photons appeared. According to this mechanism, photons were the last to appear. Moreover, the total energy of photons generated in this way should be much smaller than the amount of matter. Taking hydrogen atoms as an example, a photon with the highest energy that can be released by the spontaneous transition is 13.6 eV, while the mass of the hydrogen atom itself is 9.38 MeV. So, their ratio is about $10^{-5}$. We have mentioned ablve that the ratio of the γ photon when a neutron and a proton combine to the mass a neutron was about $10^{-3}$. Of course, the ratio of the energy level difference inside a system and the mass of the system itself differs from system to system. Generally speaking, this ratio should be relatively small. Qualitatively, there should be $\Omega_{M0} > \Omega_{R0}$.

In summary, based on our analysis, it is conjectured that the order of appearance of the four ingredients in the universe is dark energy $\Omega_{\Lambda 0} \to$ dark matter $\Omega_{DM0} \to$ matter $\Omega_{M0} \to$ photons $\Omega_{R0}$. Their relative proportions are

$$\Omega_{\Lambda 0} > \Omega_{DM0} > \Omega_{M0} > \Omega_{R0}. \tag{83}$$

This relationship is in agreement with Eq. (1). Thus, we can say that we qualitatively explain Eq. (1). At present, we are unable to give quantitative calculation for the four ingredients. What we are talking about here is the behavior of the universe at very early time.

## VII. CONCLUSION

In this paper, we have made further research on the topics related to NKE substances based on the series of work done previously. The collisions of a NKE particle and a PKE particle are investigated. A possible way to accelerate particles is suggested.

We propose the Cosmic Dark Radiation Background and Gravity Potential Background. The total energy of our universe should be negative.

We put forth a mechanism by which NKE substances can constitute a stable PKE system in terms of attractions between them. Two macroscopic NKE objects can compose a stable PKE system by gravity. Two microscopic particles, if there is attraction between them, can also form a stable PKE system. According to this mechanism, all NKE matter in the universe can form stable PKE systems. If two macroscopic PKE objects compose a stable system by gravity, they must collide with a third party to release part of their PKE. Therefore, if there are only PKE objects in the universe, there will always be some objects that cannot participate in stable PKE systems. Nevertheless, in our observable universe, it seems that all celestial bodies constitute stable PKE systems, although in different structural levels.

According to the mechanism mentioned above, starting from microscopic particles, NKE substances can constitute larger and larger stable PKE systems. Only the centroid of the highest structure has a NKE. We are unble to probe whether the center of mass of a celestial structure like a galaxy is of PKE or NKE.

It is currently believed that there are four ingredients in the universe, photons $\Omega_{R0}$, matter $\Omega_{M0}$, dark matter $\Omega_{DM}$, and dark energy $\Omega_{\Lambda0}$. We believe that the universe was originally filled with dark energy at very early time. The other three ingredients were produced in the order as follows: dark energy → dark matter → matter → photons. According to our analysis, the proportions of these four components should be qualitatively $\Omega_{\Lambda0} > \Omega_{DM0} > \Omega_{M0} > \Omega_{R0}$.

Our whole theory is based on the fact that the negative energy solutions of the Dirac equation should be interpreted as particles with NKE, which are dark particles. Negative kinetic energy systems naturally have negative pressure and negative temperature. We do not need to come up with any new particles and related new concepts.

**ACKNOWLEDGMENTS**
This work is supported by the National Natural Science Foundation of China under Grant No. 12234013, and the National Key Research and Development Program of China Nos. 2018YFB0704304 and 2016YFB0700102.